\documentclass[multphys,vecarrow]{svmult}

\usepackage{graphicx}
\usepackage{amssymb}
\usepackage{amsmath}
\usepackage{subeqnar}
\usepackage{epsfig}
\usepackage{citesort}
\begin{document}
\frontmatter%%%%%%%%%%%%%%%%%%%%%%%%%%%%%%%%%%%%%%%%%%%%%%%%%%%%%%
%\input{BOOK/titlepage}
%\input{BOOK/preface}
%\tableofcontents
\mainmatter%%%%%%%%%%%%%%%%%%%%%%%%%%%%%%%%%%%%%%%%%%%%%%%%%%%%%%%

\bibliographystyle{unsrt}

\title*{Production, Supply, and Traffic Systems:\\ A Unified Description}
\author{Dirk Helbing}
\institute{
Institute for Economics and Traffic,
Dresden University of Technology\\
Andreas-Schubert-Str. 23, 01062 Dresden, Germany, {\tt www.helbing.org}\\
e-mail: {\tt helbing@trafficforum.org}\\
}
\authorrunning{D. Helbing}

\date{\today}

\maketitle

\begin{abstract}
The transport of products between different suppliers or production units can be 
described similarly to driven many-particle and traffic systems. We introduce
equations for the flow of goods in supply networks and the adaptation of
production speeds. Moreover, we present two examples: The case of
linear (sequential) supply chains and the case of re-entrant production. In particular,
we discuss the stability conditions, dynamic solutions, and resonance phenomena
causing the frequently observed ``bullwhip effect'', which is an analogue of stop-and-go
traffic. Finally, we show how to treat discrete units and cycle times, which can be 
applied to the description of vehicle queues and travel times in freeway networks.\\[2mm]
\noindent
{PACS: 89.40.+k,89.65.Gh,47.70.-n,89.75.Hc,89.20.Bb}\\[2mm]
{\bf Keywords:} Stop-and-go traffic, supply network, production system, bullwhip effect, resonance, 
convective instability, re-entrant production, cycle time, travel time, queueing network\\
\end{abstract}

Supply chain management is a major subject in economics, as it significantly 
determines the efficiency of production processes \cite{FactPhys}. Many related studies focus on
subjects like optimum buffer sizes and stock levels, but a stable dynamics and
optimal network structure are also important subjects \cite{stabilization,prl}.
Therefore, this scientific field is connected with the statistical physics of
networks \cite{nets1}. 
\par
In this paper, we will investigate the dynamic properties and linear stability of
supply chains by a generalization of ``fluid-dynamic'' production models, which
have been inspired by traffic models \cite{Witt,Daganzo,Arm,radons,Lefeber,business}. Fluid-dynamic
models take into account non-linear interactions and are suitable for on-line control,
as they are numerically much more efficient than event-driven (Monte-Carlo) simulations.
Moreover, unlike queueing theory, they are not mainly restricted to the 
treatment of stationary situations, but can reflect variations in the consumption rate and
effects of machine breakdowns or changes in the production schedule. 
\par
The organization of this contribution is as follows: Section~\ref{Sec1} focusses on the description
of supply and production systems, in particular linear supply chains and re-entrant production.
Section~\ref{Sec2} treats freeway traffic as a queueing network and applies the formula for
the cycle time to the determination of travel times.

\section{Modeling production systems as supply networks} \label{Sec1}

Our  production model assumes $u$ production units or suppliers $j$ which deliver
$d_{ij}$ products of kind $i\in \{1,\dots,p\}$ per production cycle to other suppliers and consume
$c_{kj}$ goods of kind $k$ per production cycle. The coefficients $c_{kj}$ and $d_{ij}$ are
determined by the respective production process, and the number of production cycles per
unit time (e.g. per day) is given by the production speed $Q_j(t)$. That is, supplier $j$ requires an average time 
interval of $1/Q_j(t)$ to produce or deliver $d_{ij}$ units of good $i$. The temporal change in the number 
$N_i(t)$ of goods of kind $i$ available in the system
is given by the difference between the inflow 
\begin{equation}
Q_i^{\rm in}(t) = \sum_{j=0}^{u} d_{ij} Q_j(t)
\end{equation} 
and the outflow 
\begin{equation}
 Q_i^{\rm out}(t) = \sum_{j=1}^{u+1} c_{ij} Q_j(t) \, .
\end{equation}
In other words, it is determined by the overall production rates $d_{ij}Q_j(t)$ of all suppliers $j$ minus their
overall consumption rates $c_{ij}Q_j(t)$:
\begin{equation}
 \frac{dN_i}{dt} = Q_i^{\rm in}(t) - Q_i^{\rm out}(t) 
 = \sum_{j=1}^u (d_{ij} - c_{ij}) Q_j(t) - Y_i(t) \, .
\label{conserv}
\end{equation}
Herein, the quantity
\begin{equation}
 Y_i(t) = \underbrace{c_{i,u+1}Q_{u+1}(t)}_{\rm consumption\ and\ losses} \quad
 - \underbrace{d_{i0} Q_0(t)}_{\rm inflow\ of\ resources}  
\end{equation}
comprises the consumption rate of goods $i$, losses, and waste (the ``export'' of material),
minus the inflows into the considered system (the ``imports''). In the following, we will assume that 
the quantities are measured in a way that $0 \le c_{ij}, d_{ij} \le 1$  (for $1 \le i \le p$, $1 \le j \le u$) 
and the ``normalization conditions''
\begin{equation}
  d_{i0} = 1 - \sum_{j=1}^u d_{ij}  \ge 0 \, , \qquad c_{i,u+1} = 1 - \sum_{j=1}^u c_{ij} \ge 0 
\end{equation}
are fulfilled. Equations (\ref{conserv}) can then be interpreted as conservation equations for the flows of goods.

\subsection{Adaptation of production speeds}

The production speeds $Q_j(t)$ may be changed in time in order to adapt to varying consumption
rates $Y_i(t)$. We will assume that it takes a typical time period $T$ to adapt the actual production speeds
$Q_j(t)$ to the desired production ones, $W_j$:
\begin{equation}
 \frac{dQ_j}{dt} = \frac{1}{T} \Big[ W_j(\{N_i(t)\},\{dN_i/dt\},\{Q_k(t)\}) - Q_j(t) \Big] \, .
\label{stra}
\end{equation}
Here, the curly brackets indicate that the so-called management or control function
$W_j(\dots)$ may depend on all inventories $N_i(t)$ with $i\in \{1,\dots,p\}$, their
derivatives $dN_i/dt$, and/or all production speeds $Q_k(t)$ with $k \in \{1,\dots,u\}$.
The resulting dynamics
of the supply network can be investigated by means of a linear stability analysis, which has been
carried out for linear supply chains in Ref.~\cite{stabilization} and for supply networks with $d_{ij} = \delta_{ij}$ in
Ref.~\cite{prl}. Some of the main results will be discussed in the following.

\subsection{Modelling one-dimensional supply chains} \label{Mod}

For simplicity, let us investigate a model of one-dimensional supply chains, here, which
corresponds to $d_{ij} = \delta_{ij}$ and $c_{ij} = \delta_{i+1,j}$, 
where $\delta_{ij}=1$ for $i=j$ and $\delta_{ij}=0$ otherwise. This implies $Q_i^{\rm in}(t)
= Q_{i}(t)$ and $Q_i^{\rm out}(t) = Q_{i+1}(t) = Q_{i+1}^{\rm in}(t)$.
The assumed model consists of a series of $u$ suppliers $i$, which 
receive products of kind $i-1$ from the next ``upstream'' supplier $i-1$ and generate products of kind $i$
for the next ``downstream'' supplier $i+1$ at a rate $Q_i(t)$ \cite{Lefeber,Arm}. 
The final products are delivered at the rate $Q_u(t)$ and removed from the system with the
consumption rate $Y_u(t) = Q_{u+1}(t)$. The consumption rate is typically subject to perturbations,
which may cause a ``bullwhip effect'' \cite{Levi,Lee1,Lee,Metters,Dej02,Dej03,Hoberg1,Hoberg2}, i.e.
growing variations in the stock levels and deliveries of upstream suppliers. This is due to delays in
the adaptation of their delivery rates. 
\par
The inventory of goods of kind $i$ changes in time $t$ according to 
\begin{equation}
\frac{dN_i }{dt} = Q_i^{\rm in}(t) - Q_i^{\rm out}(t) = Q_i (t) - Q_{i + 1} (t) \, ,
\label{eq3}
\end{equation}
while the temporal change of the delivery rate will be assumed as 
\begin{equation}
 \frac{dQ_i}{dt} = \frac{1}{T} \Big[ W_i(N_i(t),dN_i/dt,Q_i(t),Q_{i+1}(t)) - Q_i(t) \Big] \, .
\end{equation}
The management strategy 
\begin{equation}
\frac{dQ_i}{dt} = \frac{1}{T} \bigg\{ \frac{N_i^0 - N_i(t)}{\tau} - \beta \frac{dN_i}{dt}  
 + \epsilon [Q_i^0 - Q_i(t)] \bigg\} 
 \label{eq4}
\end{equation}
appears to be appropriate to keep the inventories $N_i(t)$ stationary, to
maintain a certain optimal inventory $N_i^0$ (in order to cope with stochastic variations due to
machine breakdowns etc.), and to operate with the equilibrium production rates $Q_i^0 = Y_u^0$,
where $Y_u^0$ denotes the average consumption rate.
The supplier-independent parameter values $\tau$, $\beta$, and $\epsilon$ can be justified with
suitable scaling arguments \cite{prl}.

\subsection{Dynamic solution and resonance effects}\label{Resonance}

In the vicinity of the stationary state characterized by $N_i(t) = N_i^0$ and $Q_i(t) = Q_i^0$, 
it is possible to calculate the dynamic solution of the one-dimensional supply chain model 
\cite{prl,radons}. For this, let $n_i(t) = N_i(t) - N_i^0$ be the deviation
of the inventory from the stationary one, and $q_i(t) = Q_i(t) - Q_i^0$ the
deviation of the delivery rate. The linearized model equations read
\begin{equation}
\label{Eq3}
\frac{dn_i }{dt} = q_i (t) - q_{i + 1} (t) 
\end{equation}
and
\begin{equation}
\label{Eq4}
\frac{dq_i }{dt} = -\frac{1}{T} \left( \frac{n_i}{\tau} + \beta \frac{dn_i}{dt} + \epsilon q_i \right) \, .
\end{equation}
Deriving Eq.~(\ref{Eq4}) with respect to $t$ and inserting Eq.~(\ref{Eq3}) results in the
following set of second-order differential equations:
\begin{equation}
\frac{d^{2}q_i}{dt^{2}}+\underbrace{\frac{(\beta+\epsilon)}{T}}_{=2\gamma}
\frac{dq_i}{dt}+\underbrace{\frac{1}{T\tau}}_{=\omega^{2}}
q_i(t) =\underbrace{\frac{1}{T} \left[ \frac{q_{i+1}(t)}{\tau} + \beta \frac{dq_{i+1}}{dt} \right]}_{= f_i(t)} \, .
\label{set}
\end{equation}
This corresponds to the differential equation for the damped harmonic oscillator
with damping constant $\gamma$, eigenfrequency $\omega$, and driving term $f_i(t)$.
The two eigenvalues of this system of equations are
\begin{equation}
\lambda_{1,2}=-\gamma\pm\sqrt{\gamma^{2}-\omega^{2}} 
= - \frac{1}{2T} \Bigg[ (\beta+\epsilon) \mp \sqrt{(\beta+\epsilon)^2 - 4T/\tau} \Bigg] \, ,
\end{equation}
i.e. for $(\beta+\epsilon) > 0$ their real parts are always negative, corresponding to a stable behavior
in time. Nevertheless, we will identify a convective instability below, i.e. the oscillation amplitude
can grow from one supplier to the next one upstream.
\par
The set of equations (\ref{set}) can be solved successively, starting with $i=u$ and
progressing to lower values of $i$. For example, assuming
periodic oscillations of the form $f_{u}(t)=f_{u}^{0}\cos(\alpha t)$, 
after a transient time much longer than $1/\gamma$ we find 
\begin{equation}
q_{u}(t)=f_{u}^{0}F\cos(\alpha t +\varphi)
\label{in}
\end{equation}
with
\begin{equation}
\tan \varphi =\frac{2\gamma\alpha}{\alpha^{2}-\omega^{2}} 
= \frac{\alpha(\beta+\epsilon)}{\alpha^2 T - 1/\tau}
\end{equation}
and 
\begin{equation}
F= \frac{1}{\sqrt{(\alpha^{2}-\omega^{2})^{2}+4\gamma^{2}\alpha^{2}}} 
= \frac{T}{\sqrt{[\alpha^{2}T-1/\tau]^2 + \alpha^2(\beta+\epsilon)^2}} \, ,
\end{equation}
where the dependence on the eigenfrequency $\omega$ is important to understand the
occuring resonance effect. Equations (\ref{set}) and (\ref{in}) imply
\begin{equation}\label{peitscheneffekt1}
f_{u-1}(t)=\frac{1}{T}\left[\frac{q_{u}(t)}{\tau} + \beta \frac{dq_{u}}{dt}\right] 
= f_{u-1}^{0}\cos(\alpha t + \varphi + \delta)
\end{equation}
with\\[-4mm]
\begin{equation}
\tan \delta =\alpha \beta \tau
\qquad \mbox{and} \qquad
f_{u-1}^{0}= f_{u}^{0} \frac{F}{T} \sqrt{(1/\tau)^2+(\alpha\beta)^{2}} \, .
\end{equation}

\subsection{``Bull-whip effect'' and stop-and-go traffic}\label{Stop}

The oscillation amplitude increases from one supplier to the next upstream one, if 
\begin{equation}
\frac{f_{u-1}^0}{f_u^0} = \bigg\{ 1 + \frac{\alpha^2[\epsilon (\epsilon + 2\beta) - 2 T/\tau ] + \alpha^4T^2 }
{(1/\tau)^2 + (\alpha\beta)^2} \bigg\}^{-1/2} > 1 \, .
\end{equation}
One can see that this resonance effect can occur for $0 < \alpha^2 < 2/(T\tau) - \epsilon(\epsilon + 2\beta)/T^2$. 
Therefore, variations in the consumption rate are magnified under the instability condition
\begin{equation}
T  > \epsilon \tau \left( \beta + \epsilon/2 \right) \, .
\label{eq15}
\end{equation}
Supply chains show this ``bullwhip effect'' (which corresponds to the phenomenon of convective, i.e. 
upstream moving instability),
if the adaptation time $T$ is too large, if there is no adaptation to the equilibrium production speed $Q_i^0$, 
corresponding to $\epsilon = 0$, or if the management
reacts too strong to deviations of the actual stock level $N_i$ from the desired one $N_i^0$,
corresponding to a small value of $\tau$.
The latter is very surprising, as it implies that the strategy
\begin{equation}
 \frac{dQ_i}{dt} = \frac{1}{T\tau} [ N_i^0 - N_i(t) ] \, ,
\end{equation}
which tries to maintain a constant work in progress $N_i(t)=N_i^0$, would ultimately lead to an 
undesireable bullwhip effect. In contrast, the management strategy
\begin{equation}
 \frac{dQ_i}{dt} = \frac{1}{T} \bigg\{ - \beta \frac{dN_i}{dt} + \epsilon [Q_i^0 - Q_i(t)] \bigg\} 
\end{equation}
would avoid this problem, but it would not maintain a constant work in progress. The 
control strategy (\ref{eq4}) with a sufficiently large value of $\tau$ would fulfill both requirements.
\par
The ``bullwhip effect'' has, for example, been reported
for beer distribution \cite{beer1,beer2}, but similar dynamical effects
are also known for other distribution or transportation chains with significant adaptation times $T$.
It has, for example, some analogy to stop-and-go traffic \cite{business,prl}, where delayed adaptation
also leads to an unstable behavior. 
In the case $\beta = 0$ and $\epsilon = 1$, the stability condition (\ref{eq15}) agrees exactly
with the one of the optimal velocity model \cite{optimvel}, which is a particular microscopic traffic
model. This car-following model assumes an acceleration equation of the form
\begin{equation}
 \frac{dv_i(t)}{dt} = \frac{V_{\rm opt}(d_i(t)) - v_i(t)}{T} 
\label{velad}
\end{equation}
and the complementary equation
\begin{equation}
 \frac{dd_i(t)}{dt} = - [v_{i}(t) - v_{i+1}(t)] \, .
\end{equation}
In contrast to the above supply chain model, however, the index
$i$ represents single vehicles, $v_i(t)$ is their actual velocity of motion, $V_{\rm opt}$ the so-called
optimal (safe) velocity, which depends on the distance $d_i(t)$ to the next vehicle ahead. 
$T$ denotes again an adaptation time. 
Comparing this equation with Eq.~(\ref{eq4}), the velocities $v_i$
would correspond to the delivery rates $Q_i$, the optimal velocity $V_{\rm opt}$ to the
desired delivery rate $W_i(N_i) = Q_i^0 + (N_i^0 - N_i)/\tau$, 
and the inverse vehicle distance $1/d_i$ would approximately
correspond to the stock level $N_i$ (apart from a proportionality factor). 
This shows that the analogy between supply chain and traffic
models concerns only their mathematical structure, but not their interpretation, although both
relate to transport processes. Nevertheless, this mathematical relationship can give us hints,
how methods, which have been successfully applied to the investigation of traffic models before,
can be generalized for the study of supply networks. 
Compared to traffic dynamics, supply networks and
production systems have some interesting new features: Instead of a continuous space,
we have discrete production units $j$, and the management strategy (\ref{stra}) is generally
different from the velocity adaptation (\ref{velad}) in traffic. With suitable strategies,
in particular with large values of $\tau$ and $\beta$,
the oscillations can be mitigated or even suppressed. 
Moreover, production systems are frequently supply networks with complex topologies rather
than one-dimensional supply chains, i.e. they have additional features compared
to (more or less) one-dimensional freeway traffic. They are more comparable to street networks of cities \cite{travel}.

\subsection{Calculation of the cycle times}

Apart from the productivity or throughput $Q_i$ of a production unit, production managers are
highly interested in the cycle time $T_i$, i.e. the time interval between the beginning of the generation
of a product and its completion. 
Let us assume that the queue length $L_i(t)$ of 
products waiting to be processed by production unit $i$
is given by the inventory $N_{i-1}(t)$ of product $i-1$. 
The change of the queue length $L_i(t)$ in time
is then determined by the difference between the arrival rate $Q_i^{\rm arr}(t)$, which
corresponds to the inflow $Q_{i-1}^{\rm in} = Q_{i-1}(t)$ from production
unit $i-1$, and the departure rate $Q_i^{\rm dep}(t)$, which corresponds to the
outflow $Q_{i-1}^{\rm out}(t) = Q_i(t)$ to production unit $i$: 
\begin{equation}
 \frac{dL_i}{dt} = \frac{dN_{i-1}}{dt} = Q_{i-1}(t) - Q_i(t) 
= Q_i^{\rm arr}(t) - Q_i^{\rm dep}(t) \, .
\label{length}
\end{equation}
On the other hand, the waiting products move forward $Q_i^{\rm dep}(t) = Q_i(t)$ 
steps per unit time, as $Q_i(t)$ is the processing rate (production speed). For this reason,
the overall time $T_i(t)$ until having been processed is given by the implicit equation
\begin{equation}
 N_{i-1}(t) = L_i(t) = \int\limits_t^{t+T_i(t)} \!\! dt' \, Q_i^{\rm dep}(t') 
=  \int\limits_{-\infty}^{t+T_i(t)} \!\! dt' \, Q_i^{\rm dep}(t')  - \int\limits_{-\infty}^{t} \!\! dt' \, Q_i^{\rm dep}(t') \, ,
\label{length2}
\end{equation}
if the queue of length $L_i(t)$ was joined at time $t$.
Accordingly, high inventories imply long cycle times, which favours just-in-time production with
small stock levels. From Eqs.~(\ref{length}) and (\ref{length2}), one can finally derive
a delay-differential equation for the waiting time under varying production conditions \cite{business}: 
\begin{equation}
 \frac{dT_i}{dt} = \frac{Q_{i}^{\rm arr}(t)}{Q_i^{\rm dep}(t+T_i(t))} - 1
= \frac{Q_{i-1}(t)}{Q_i(t+T_i(t))} - 1 \, .
\label{tim}
\end{equation}
This equation can be solved numerically as a function of the production rates
$Q_i(t)$, since the production initially starts with a cycle time of $T_i(0) = 1/Q_i(0)$, corresponding to the
average processing time $1/Q_i(0)$ when the production
unit $i$ is started. In this way, it is possible to determine the process cycle times $T_i$ and the overall
production time as the sum of the cycle times of all single production steps, taking into account
the respective time delays $T_i$: When a specific product enters the queue before production unit $i$ at time
$t=t_{i}$, it enters the queue before production unit $i+1$ at time $t_{i+1} = t_i + T_i(t_i)$. The
time of delivery to the customer is given by $t_{u+1} = t_u + T_u(t_u)$. 

\subsection{Modeling of discrete units}

The above ``fluid-dynamic'' model equations can not only be used to represent approximate
mean values of large numbers of products. They can also be transfered to the treatment of discrete units 
(such as single units of a product), if their dynamics is sufficiently deterministic.
For example, one could represent the time interval $\Delta T$, during which a discrete
unit occupies a certain production unit $i$ by a step function. However, as step functions are not
differentiable everywhere, we will replace them by smooth functions.
\par
One possible specification uses a Fourier approximation of the step functions. To lowest order, one may take
\begin{equation}
 Q_i(t) = A \sum_k \{1 - \cos[\pi (t-t_i^k)/\Delta T ] \} 
\label{Q1}
\end{equation}
where $t_i^k$ denotes the starting time of occupation by object $k$. The cosinus function
is set to zero for $t< t_i^k$ and $t> t_i^k+\Delta T$. 
The prefactor $A$ is determined in a way that satisfies the normalization condition,\\[-2mm]
\begin{equation}
 \int\limits_{t_i^k}^{t_i^k+\Delta T} \!\! dt' \, Q_i(t') = 1 \, , \\[-1.5mm]
\label{norma}
\end{equation}
which implies $A = 1/\Delta T$. Under this condition, the 
integral 
\begin{equation}
N_i^{\rm tot}(t) = \int_0^t dt' \, Q_i(t')
\label{total}
\end{equation}
counts the number of units that have passed the cross section or production unit under consideration.
\par
Another possible specification assumes 
\begin{equation}
 Q_i(t) = B \sum_k (t-t_i^k)^2 (t - t_i^k - \Delta T)^2 \, ,
\label{Q2}
\end{equation}
where we set the fourth order polynomial to zero for $t< t_i^k$ and $t> t_i^k+\Delta T$. 
The normalization condition (\ref{norma}) implies $B=30/(\Delta T)^5$. 
\par
Together with (\ref{Q1}) or (\ref{Q2}),
the relationship (\ref{total}) can also be applied to situations where several units occupy a 
production unit at the same time. As before, the management function can be chosen as
a function of the inventories $N_i(t)$, but in many cases, 
it would be reasonable to replace a reaction to temporal changes $dN_i/dt$ in the inventories 
by exponentially smoothed values.

\subsection{Re-entrant production}

\begin{figure}
\vspace*{-5mm}
\centerline{\epsfig{file=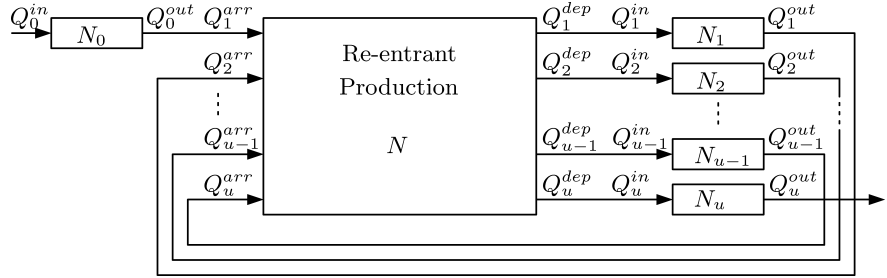}} 
\label{reentrant}
\caption{Illustration of the model of re-entrant production and its variables (see main text for details).}
\vspace*{-3mm}
\end{figure}
Semiconductor production \cite{Arm1} relies on some extremely expensive (lithographic) production units,
which are therefore used for many similar subsequent production steps $i$, namely the production
of the various layers of a chip. This is the reason for re-entrant production \cite{reent1,reent2} 
(see Fig.~\ref{reentrant}), posing particular control problems not only for the overall arrival rate $Q^{\rm arr}(t)$ in
the re-entrant production area, but also concerning the fractions $p_i(t)$ of  products 
in the different production stages $i$ fed into it. If $N_i(t)$ denotes the stock level of
chips after the $i$th entry (production step), $Q_i^{\rm in}(t)$ the respective inflow, and 
$Q_i^{\rm out}(t)$ the outflow, the related balance equation is again
\begin{equation}
 \frac{dN_i}{dt} = Q_i^{\rm in}(t) - Q_i^{\rm out}(t) \, .
\end{equation}
Assuming also a buffer before the first and a buffer after the last ($u$th) production step, 
this equation applies to $i\in\{0,1,\dots,u\}$, where $Q_{u}^{\rm out}$ corresponds to
the removal rate of products that have completed the re-entrant (lithographic) production steps.
The outflow $Q_{i-1}^{\rm out}(t)$ determines the arrival rate $Q_i^{\rm arr}(t)$
for the $i$th re-entrant production step, while the departure rate $Q_i^{\rm dep}(t)$ determines
the inflow $Q_i^{\rm in}(t)$ into the subsequent buffer. The overall stock level $N(t)$ of 
chips at various production stages in the re-entrant production area changes in time according to 
\begin{equation}
 \frac{dN}{dt} = \sum_{i=1}^{u} [Q_{i-1}^{\rm out}(t) - Q_i^{\rm in}(t)] 
 = \sum_{i=1}^{u} [Q_i^{\rm arr}(t) - Q_i^{\rm dep}(t)] = Q^{\rm arr}(t) - Q^{\rm dep}(t) \, .
\end{equation}
The overall departure rate $Q^{\rm dep}(t)$ from the re-entrant production area depends on 
the number $N(t)$ of products processed at the same time. Moreover, if $p_i(t)$ denotes the
fraction of chips that have entered the re-entrant area for the $i$th production step at time $t$
and $T^0(t)$ is the related cycle time of these chips in the re-entrant area, their departure
rate is given by 
\begin{equation}
 Q_i^{\rm dep}(t) = p_i(t-T^0(t)) Q^{\rm dep}(N(t)) = Q_i^{\rm in}(t) \, .
\end{equation}
As before, the delay-differential equation for the temporal change of the cycle time is 
\begin{equation}
 \frac{dT^0}{dt} = \frac{Q^{\rm arr}(t)}{Q^{\rm dep}(t+T^0(t))} -1 \, .
\end{equation}
The overall arrival rate may be adapted according to 
\begin{equation}
 \frac{dQ^{\rm arr}}{dt} = \frac{1}{T} [W(N(t),dN/dt,Q^{\rm arr}(t),Q^{\rm dep}(t),N_{u}(t), dN_{u}/dt, Q_{u}^{\rm out}(t)) 
- Q^{\rm arr}(t)] \, ,
\end{equation}
where the management function $W(\dots)$ depends not only on the variables $N(t)$ and
$Q^{\rm dep}(t)$ characterizing the re-entrant production area, but also on the (desired)
removal rate $Q_{u}^{\rm out}(t)$ and the stock level $N_{u}(t)$ of the final buffer.
Finally, the specific arrival rates of chips for the $(i+1)$st re-entrant production step are
\begin{equation}
 Q_{i+1}^{\rm arr}(t) = p_{i+1}(t) Q^{\rm arr}(t) = Q_i^{\rm out}(t) \, ,
\end{equation}
where their relative percentages
\begin{equation}
 p_{i+1}(t) = \frac{W_i(N_i(t),dN_i/dt,Q_i^{\rm in}(t),Q_i^{\rm out}(t),Q_{u}^{\rm out}(t))}
{\sum_j W_j(N_j(t),dN_j/dt,Q_j^{\rm in}(t),Q_j^{\rm out}(t),Q_{u}^{\rm out}(t))}
\end{equation}
are controlled by another management function $W_i(\dots)$, which depends on the
stock levels $N_i(t)$ of the respectively preceding buffers and on the removal rate $Q_{u}^{\rm out}(t)$.
The management functions $W_i(\dots)$ allow one to specify different priorities such as push or
pull strategies. Specifications of the management function, which can avoid bullwhip (resonance) effects
and long cycle times will be presented in a forth-coming paper. Generalizations to the simultaneous
production of various products are possible, requiring the consideration of
an individual number of re-entrant steps with potentially
different production speeds and cycle times (``overtaking'')
in the re-entrant area. The problem
is similar to the treatment of heterogeneous multi-lane traffic \cite{multi}. In the following, we will present
a simpler traffic model for uniform vehicles, which shows how to transfer the above approach
from production processes to transportation processes in street networks.

\section{Queueing model of vehicle traffic in freeway networks} \label{Sec2}

In the following, we will propose a traffic model \cite{travel}, which
was inspired by the above model of supply networks.
Here, the elements to be served are the vehicles. The formula for the 
queue length determines the density in a road section, and the cycle time corresponds
to the travel time. 
\par
When we now specify road traffic as a queueing system, we will take into account essential 
traffic characteristics such as the flow-density relation or the properties of extended congestion patterns 
at bottlenecks. In fact, traffic congestion is usually triggered by spatial
inhomogeneities of the road network \cite{phase}, 
and queueing effects are normally not observed along sections of low
capacity, but upstream of the beginning of a bottleneck. Therefore, 
we will subdivide roads into sections $i$ of homogeneous capacity and length $l^{\rm max}_i$,
which start at place $x_i$ and
end with some kind of inhomogeneity (i.e. an increase or decrease of capacity) at place $x_{i+1}= x_i+l^{\rm max}_i$.
In other words, the end of a road section $i$ is, for example, determined by the location
of an on- or off-ramp, a change in the number of lanes, or the beginning or end of a gradient.
\par
The model can be derived from the ``fluid-dynamic'' continuity equation 
\begin{equation}
 \frac{\partial \rho(x,t)}{\partial t} + \frac{\partial Q_i(x,t)}{\partial x} = 
 \mbox{source terms} 
\end{equation}
describing the conservation of the number of vehicles.
Here, $\rho(x,t)$ denotes the vehicle density per lane at place $x$ and time $t$, and
$Q_i(x,t)$ the traffic flow per lane. The source terms originate from ramp flows
$Q_i^{\rm ramp}(t)$, which enter the road at place $x_{i+1}$. Let us define the
arrival rate at the upstream end of road section $i$ (the ``inflow'') by $Q_i^{\rm arr}(t)= I_i Q_i(x_i+dx,t)$,
where $dx$ is a differential space interval and $I_i$ the number of lanes of road
section $i$. Analogously, the departure rate from the downstream end of this section is defined by
$Q_i^{\rm dep}(t) = I_i Q_i(x_{i+1}-dx,t)$. (Note that this ``outflow'' from section $i$ is to be distinguished from the 
outflow $Q_{\rm out}$ per lane from congested traffic).
The conservation of the number of vehicles implies that the departure rate plus the ramp flow determine
the arrival rate in the next downstream section $i+1$:
\begin{equation}
 Q_{i+1}^{\rm arr}(t) = Q_i^{\rm dep}(t) + Q_i^{\rm ramp}(t) \, .
\label{arr}
\end{equation}
In order to guarantee non-negative flows, we will demand for the ramp flows that the consistency condition
$- Q_i^{\rm dep}(t) \le Q_i^{\rm ramp}(t) \le Q_{i+1}^{\rm arr}(t)$ is always met.
\par
Integrating the continuity equation over $x$ with $x_i < x < x_{i+1}$ provides a conservation
equation for the number $N_i(t)
= \int_{x_i}^{x_{i+1}} dx \; I_i \rho(x,t)$ of vehicles in road section $i$. It changes according to 
\begin{equation}
 \frac{dN_i(t)}{dt} = Q_i^{\rm arr}(t) - Q_i^{\rm dep}(t) = Q_{i-1}^{\rm dep}(t) + Q_{i-1}^{\rm ramp}(t)  
 - Q_{i}^{\rm dep}(t) \, .
\label{length3}
\end{equation}
In the terminology of queueing theory,
equation (\ref{length3}) reflects the change in the number $N_i(t)$ of vehicles
waiting to be served by the downstream end of the section, and 
the number $I_i$ of lanes corresponds to the number of ``channels'' serving in parallel. Moreover, 
$N_i(t)$ corresponds to the ``queue length''. However, this does not necessarily mean that we have congested traffic
as, in the terminology of traffic theory, ``queue length'' refers to something else, namely the spatial extension 
$l_i(t) > 0$ of a  congested road section. 
In the following, we will try to express the traffic dynamics and the travel times only through the 
flows at the cross sections $x_i$, taking into account the features
of traffic flow in a simplified way.
\par
For free flow, i.e. below some critical vehicle density 
$\rho_{\rm cr}$ per lane, the relation between the traffic flow $Q_i$ per lane and the density $\rho$ per lane
can be approximated by an increasing linear relationship, 
while above it, a falling linear relationship 
is consistent with congested flow-density data (in particular, if the average time gap
$T$ is treated as a time-dependent, fluctuating variable) \cite{gaps}. This implies $Q_i(x,t)
\approx Q_i(\rho(x,t))$ with
\begin{equation}
 Q_i(\rho) = \left\{
\begin{array}{ll}
Q_i^{\rm free}(\rho) = \rho V_i^0 & \mbox{if } \rho < \rho_{\rm cr} \\
Q_i^{\rm cong}(\rho) = ( 1 - \rho / \rho_{\rm jam}) / T & \mbox{otherwise.}  
\end{array}\right. 
\label{Q}
\end{equation}
Here, $V_i^0$ denotes the average free velocity,
$T$ the average time gap, and $\rho_{\rm jam}$ the density per lane
inside of traffic jams. Moreover, we define the free and congested densities by
\begin{equation}
 \rho_i^{\rm free}(Q_i) = Q_i / V_i^0 \quad \mbox{and} \quad
 \rho_i^{\rm cong}(Q_i) = (1-TQ_i)\rho_{\rm jam} \, .
\label{def}
\end{equation}
The quantity $Q_{\rm out} = (1-\rho_{\rm cr}/\rho_{\rm jam})/T$ 
corresponds to the outflow per lane from congested
traffic \cite{outflow}. Depending on the parameter specification,
the model describes a continuous flow-density relation 
(for $\rho_{\rm cr} V_i^0 = Q_{\rm out}$) or a 
capacity drop at the critical density $\rho_{\rm cr}$ and high-flow states immediately before
(if $\rho_{\rm cr} V_i^0 > Q_{\rm out}$). 
\par
According to shock wave theory \cite{LW}, density variations at place $x$ propagate with velocity
$C(t) = [Q_i(x+dx,t)-Q_i(x-dx,t)]/[\rho(x+dx,t)-\rho(x-dx,t)]$. 
Accordingly, the propagation velocity is $C=V_i^0$ in free traffic, and $C= - c = 
- 1/(T\rho_{\rm max})$ in congested traffic. Therefore, it takes the time period $T_i^{\rm free} =
l^{\rm max}_i/V_i^0$ for a perturbation to travel through free traffic, while it takes the time period $T_i^{\rm cong}
= l^{\rm max}_i/c$, when the entire road section $i$ is congested.
\par
Now, remember that congestion in section $i$ starts to form upstream of a
bottleneck, i.e. at place $x_{i+1}$. Let $l_i(t)$ denote the length of the congested area and
$x(t) = x_{i+1}-l_i(t) = x_i+l^{\rm max}_i-l_i(t)$ the location of its upstream front. Then, we have free traffic
between $x_i$ and $x_i+l^{\rm max}_i-l_i(t)$, i.e. $Q_i(x-dx,t) = Q_i^{\rm arr}(t - (x-x_i)/V_i^0)$ (considering $dx \rightarrow 0$), 
and congested traffic downstream
of $x(t)$, i.e. $Q_i(x+dx,t) = Q_i^{\rm dep}(t - (x_{i+1}-x)/c)$. With $dx/dt = -dl_i/dt = C(t)$ and Eq.~(\ref{def}) we find
\begin{equation}
 \frac{dl_i}{dt} = -\frac{Q_i^{\rm dep}(t-l_i(t)/c)/I_i - Q_i^{\rm arr}(t-[l^{\rm max}_i-l_i(t)]/V_i^0)/I_i}
{\rho_i^{\rm cong}(Q_i^{\rm dep}(t-l_i(t)/c)/I_i) - \rho_i^{\rm free}(Q_i^{\rm arr}(t-[l^{\rm max}_i-l_i(t)]/V_i^0)/I_i)} \, .
\label{see}
\end{equation}
\par
The capacity of a {\em congested} road section $i$ is approximated as the outflow 
$Q_{\rm out} = (1-\rho_{\rm cr}/\rho_{\rm jam})/T$ from congested traffic per lane
times the number $I_i$ of lanes, minus the maximum bottleneck strength at the end of this section.
This may be given by an on-ramp flow $Q_i^{\rm ramp}(t) > 0$ or analogously by $(I_{i} - I_{i+1})Q_{\rm out}$ 
in case of a reduction $I_{i+1}-I_i < 0$ in the number of lanes, or in general by some time-dependent value
$\Delta Q_i(t)$ in case of another bottleneck such as a gradient:
\begin{equation}
 Q_i^{\rm cap}(t) = I_i Q_{\rm out} - \max[ Q_i^{\rm ramp}(t), (I_{i}-I_{i+1})Q_{\rm out}, \Delta Q_i(t), 0] \, .
\end{equation} 
Analogously, the maximum capacity $Q_i^{\rm max}(t)$ of the road section $i$ under free flow conditions is
given by the maximum flow $I_i \rho_{\rm cr} V_i^0$ minus the reduction by bottleneck effects:
\begin{equation}
 Q_i^{\rm max}(t) = I_i \rho_{\rm cr} V_i^0 - \max[ Q_i^{\rm ramp}(t), (I_{i}-I_{i+1})\rho_{\rm cr} V_i^0, \Delta Q_i(t), 0] \, .
\end{equation} 
\par
Moreover, we have to specify the departure rate $Q_i^{\rm dep}(t)$
as a function of the respective traffic situation. 
Focussing on the cross section at location $x_{i+1}$
and considering the directions of information flow (i.e. the propagation direction of density variations),
we can distinguish three different cases:
\begin{itemize}
\item[1.] If we have free traffic in the upstream section $i$ and free or partially congested
traffic in the downstream section $i+1$, density variations propagate downstream and 
the departure rate $Q_i^{\rm dep}(t)$ at time $t$ 
is given as the arrival rate $Q_i^{\rm arr}(t-T_i^{\rm free}) 
= Q_{i-1}^{\rm dep}(t-T_i^{\rm free}) + Q_{i-1}^{\rm ramp}(t-T_i^{\rm free})$,
since the vehicles entering section $i$ at time $t-T_i^{\rm free}$ 
leave the section after an average travel time $T_i$ of $T_i^{\rm free}$.
\item[2.] In the case of partially or completely congested traffic 
upstream and free or partially congested traffic downstream,
the departure rate $Q_i^{\rm dep}(t)$ is given by
the capacity $Q_i^{\rm cap}(t)$ of the congested road section $i$.  
\item[3.] In the case of congested traffic on the entire downstream road section $i+1$,
the departure rate $Q_i^{\rm dep}(t)$ is given by
the departure rate $Q_{i+1}^{\rm dep}(t-T_{i+1}^{\rm cong})$ from the downstream section 
at time $t-T_{i+1}^{\rm cong}$ minus the ramp flow $Q_i^{\rm ramp}(t)$ entering at location $x_{i+1}$.
\end{itemize}
Summarizing this, we have
\begin{equation}
 Q_i^{\rm dep}(t) = \left\{
\begin{array}{ll}
\!Q_{i}^{\rm arr}(t-T_i^{\rm free}) & \mbox{if } l_{i+1}(t) < l^{\rm max}_{i+1} \mbox{ and } l_i(t) = 0 \\
\!Q_i^{\rm cap}(t) & \mbox{if } l_{i+1}(t) < l^{\rm max}_{i+1} \mbox{ and } l_i(t) > 0 \\
\!Q_{i+1}^{\rm dep}(t-T_{i+1}^{\rm cong})-Q_i^{\rm ramp}(t) & \mbox{if } l_{i+1}(t) = l^{\rm max}_{i+1}.
\end{array}
\right.
\label{rates}
\end{equation}
\par
A numerical solution of the above defined section-based queueing-theo\-re\-ti\-cal traffic model is carried out as
follows: First, calculate the new arrival and departure 
rates by means of Eqs.~(\ref{arr}) and (\ref{rates}), taking into account
the boundary conditions for the flows at the open ends of the road network. 
Second, determine the queue lengths $l_i(t)$ in all road sections $i$: (i) If traffic in the road section 
flows freely ($l_i(t) = 0$) and the maximum capacity $Q_i^{\rm max}(t)$ is not reached,
i.e. $Q_i^{\rm arr}(t-T_i^{\rm free}) < Q_i^{\rm max}(t)$, we have $dl_i(t)/dt = 0$
and the traffic flow in the road section remains free. (ii) If the road section is
completely congested ($l_i(t) = l^{\rm max}_i$) and the arrival rate $Q_i^{\rm arr}(t)$ 
is not below the departure rate at time $t- T_i^{\rm cong}$, 
i.e. $Q_{i}^{\rm dep}(t-T_i^{\rm cong}) \le Q_{i}^{\rm arr}(t)$,
the road section $i$ stays fully congested and $dl_i/dt=0$.
(iii) In other cases, we have partially congested traffic in road section $i$ 
and the length $l_i(t)$ of the congested area changes according to Eq.~(\ref{see}).
Next, one continues with the first step for the new time $t+dt$, and so on. It is obvious, that
this numerical solution is significantly more simple and robust than the numerical solution of
the Lighthill-Whitham model, as shock waves (i.e. the interfaces between free and congested traffic)
are treated analytically and the propagation velocities of
perturbations within the free and congested regions are constant.
\par
The travel time $T_i(t)$ of a vehicle that enters road section $i$ at time $t$
can be calculated analogously to Eq.~(\ref{tim}) \cite{travel}, i.e.
via the delay-differential equation
\begin{equation}
 \frac{dT_i(t)}{dt} = \frac{Q_i^{\rm arr}(t)}{Q_i^{\rm dep}(t+T_i(t))} - 1 
 = \frac{Q_{i-1}^{\rm dep}(t) + Q_{i-1}^{\rm ramp}(t)}{Q_i^{\rm dep}(t+T_i(t))} - 1 \, .
\label{travtime}
\end{equation}
According to this, the travel time $T_i(t)$ increases with time, when the arrival rate $Q_i^{\rm arr}$ 
at the time $t$ of entry exceeds the departure rate $Q_i^{\rm dep}$ at the leaving time $t+T_i(t)$, while it decreases when it
is lower. It is remarkable that this formula does not explicitly depend on the velocities on the road section, but
only on the arrival and departure rates. The calculation of the
travel time based on the velocity is considerably more complicated: Let $v(t)$ be the velocity and 
$x(t) = x_i + \int_{t_0}^t dt' \; v(t')$ the location of a vehicle
at time $t$, when it enters section $i$ at time $t_0$. Its travel time $T_i(t_0)$ on section $i$ is given by
the implicit equation $x(t+T_i(t_0)) = x_{i+1}$, which says that this vehicle reaches
place $x_{i+1} = x_i + l^{\rm max}_i$ at time $t_0+T_i(t_0)$. The vehicle speed $v(t)$ is
also difficult to determine, as it depends on the (free or congested) traffic state
at its respective location $x(t)$: It is $v(t) = V_i^0$ in free flow, i.e. for $V_i^0 (t-t_0) < l^{\rm max}_i - l_i(t)$. 
In congested flow, i.e. for $t_0 + [l^{\rm max}_i-l_i(t)]/V_i^0 \le t \le t_0+T_i(t_0)$, it is determined via 
$v(t) = Q_i(x(t),t)/\rho(x(t),t) = [1/\rho(x(t),t) - 1/\rho_{\rm max}]/T$ with 
$\rho(x(t),t) = \rho(x_{i+1},t-[x_{i+1}-x(t)]/c) = \{1 -T Q_i^{\rm dep}(t-[x_{i+1}-x(t)]/c)/I_i\}\rho_{\rm jam}$.

\section{Summary and conclusions}

In this contribution, traffic and production systems have been treated in a uniform way as dynamic
queueing networks, since it does not matter whether one treats the transport of goods from one
production unit to the next one or of vehicles from one cross section of the road network to the next.
Consequently, the stability conditions for supply chains looked similar to the ones of some specific 
traffic models. Moreover, we have presented a delay-differential equation for the determination of 
cycle (production) times, which can be also applied to calculate the travel times of vehicles. Interestingly
enough, this formula did not require to calculate the vehicle speed 
on freeway sections, but only the in- and outflows at certain cross sections of the street network,
namely, where the road capacity changed.
\par
The derived instability conditions allow one to choose appropriate management strategies which can
avoid the well-known bullwhip effect. This effect describes an amplification of variations
in the delivery rate and inventory from one supplier to the next one upstream. Such a convective
instability can occur despite of a stable behavior in time because of the possibility of resonance effects. 
Apart from this, we have sketched the treatment of re-entrant production processes and of
discrete units.
\par
The advantage of ``fluid-dynamic'' models of traffic, supply, and production networks is their
great numerical efficiency and their consideration of non-linear interaction effects. Therefore,
in contrast to most classical queueing theoretical approaches and to event-driven (Monte Carlo)
simulations, they are suitable for on-line control and the treatment of variations in the consumption rate
or in the production program. 
\par
Present research focusses on the effect of the topology of supply networks
on their dynamics \cite{prl}. The dynamics of production processes can even be chaotic 
\cite{beer1,Chaos1,Chaos2,Chaos3}.
Other studies concentrate on the subject of optimal control (including chaos control)
\cite{Chaos3,control5,control2,control3,control4,Zipkin,control1}, which is particularly challenging for
re-entrant production \cite{reent1,reent2}. 
Finally, the presented traffic model is now being applied to the simulation of
city networks with adaptive traffic light control and to dynamic assignment problems.

\subsection*{Acknowledgements}

The author would like to thank Stefan L\"ammer, Martin Treiber, and Thomas Seidel for their
valuable comments. This work was partially supported by the German Research Foundation (DFG), grant
no. He 2789/5-1.

\end{document}